\documentclass[prb,twocolumn,nofootinbib,citeautoscript,10pt,notitlepage]{revtex4-2}
\pdfoutput=1

\usepackage{dsfont}
\usepackage{float}
\usepackage{array}
\usepackage{slashed,bbold}
\usepackage{physics}
\usepackage{enumerate}

\usepackage{amsmath,amssymb,bm} 
\usepackage{graphicx}

\usepackage[tight]{subfigure} 

\usepackage{color} 
\usepackage[papersize={8.5in,11in}]{geometry}

\usepackage{color}
\definecolor{darkblue}{rgb}{0.,0.,0.4}
\definecolor{darkred}{rgb}{0.5,0.,0.}
\definecolor{BlueViolet}{RGB}{138,43,226}
\definecolor{SkyBlue}{RGB}{30,144,255}
\definecolor{DarkGreen}{RGB}{0,100,0}
\usepackage[pdftex,colorlinks=true,linkcolor=darkblue,citecolor=blue,urlcolor=darkred]{hyperref}

\renewcommand{\epsilon}{\varepsilon}

\def \be{\begin{equation}}
\def \ee{\end{equation}}
\def \nn{\nonumber \\}
\begin{document}

\title{Andreev bound states in Josephson junctions of semi-Dirac semimetals}

\author{Ipsita Mandal}
\affiliation{Department of Physics, Shiv Nadar Institution of Eminence (SNIoE), Gautam Buddha Nagar, Uttar Pradesh 201314, India}

\begin{abstract}
We consider a Josephson junction built with the two-dimensional semi-Dirac semimetal, which features a hybrid of linear and quadratic dispersion around a nodal point. We model the weak link between the two superconducting regions by a Dirac delta potential because it mimics the thin-barrier-limit of a superconductor-barrier-superconductor configuration. Assuming a homogeneous pairing in each region, we set up the BdG formalism for electronlike and holelike quasiparticles propagating along the quadratic-in-momentum dispersion direction. This allows us to compute the discrete bound-state energy spectrum $\varepsilon $ of the subgap Andreev states localized at the junction. 
In contrast with the Josephson effect investigated for propagation along linearly dispersing directions,
we find a pair of doubly degenerate Andreev bound states. Using the dependence of $\varepsilon $ on the superconducting phase difference $\phi$, we compute the variation of Josephson current as a function of $\phi$.
\end{abstract}

\maketitle

\tableofcontents


\section{Introduction}

In a configuration consisting of two superconductors coupled together by a weak link between them, a dissipationless current flows across the junction in equilibrium \cite{josephson,likharev,waldram}, which is dubbed as the \textit{Josephson current} $I_J$.
$I_J$ is a single-valued and $2\pi$-periodic function in the phase difference $\phi$ of the pair potential of the two
superconductors. In such a set-up, the Andreev surface states of the two superconductors hybridize to form Andreev bound states (ABSs) at the junction. These states are the dominant contributors to the Josephson current through the junction \cite{been_houten,titov-graphene,zagoskin,tanaka_review}, because the contributions from the excited states in the continuum are negligible.
There is an extensive literature devoted to the study of such Josephson effects in two-dimensional (2d) and three-dimensional (3d) materials like graphene and Weyl-like semimetals \cite{titov-graphene,bolmatov_graphene_sns,krish-moitri,emil_jj_WSM,sumathi-udit,debabrata-krish,debabrata,ips_jj_rsw}, where the weak link is a tunneling barrier.
In other words, the two superconducting regions (each abbreviated by ``S'') are weakly coupled by a middle region consisting of the normal (i.e., non-superconducting) phase (abbreviated by ``N'') of a semimetal.
Two alternate configurations include the S-N-S~\cite{titov-graphene,bolmatov_graphene_sns} and the S-B-S (where ``B'' indicates a potential barrier in the N region) \cite{krish-moitri,emil_jj_WSM,debabrata-krish,debabrata,ips_jj_rsw} junctions.
While the superconductivity is induced via proximity-effect by placing a conventional s-wave superconductor on top of the corresponding electrode \cite{proximity-sc}, the tunnel barrier can be created by applying a gate voltage $V_0$ across N.
A schematic representation of the S-B-S set-up is illustrated in Fig.~\ref{figsetup}(a).

The characterization of the Josephson junctions in the Dirac/Weyl-like systems, described above, has spanned both isotropic and anisotropic bandstructures. For the anisotropic cases, in addition to a linear-in-momentum dispersion along one of the momentum axes (let us call it $k_z$), there exists quadratic(cubic)-in-momentum variations along the remaining axes orthogonal to $k_z$ \cite{debabrata-krish,debabrata}. However, in such studies, the propagation direction has always been chosen to be along the linear-in-momentum dispersion axis. To fill in this gap, we consider here a 2d semi-Dirac  semimetal, which features a quadratic dispersion along the axis perpendicular to $k_z$, which we label as the $k_x$-axis, and consider the propagation of the quasiparticles/quasiholes along the quadratic dispersion direction (i.e., along the $x$-axis).
Such hybrid dispersion characteristics appear in
the low-energy spectra of a tight-binding model on the (1) honeycomb lattice in a magnetic field (resulting in the so-called Hofstadter spectrum) \cite{dietl} and (2) square-lattice with three bands of spinless fermions \cite{banerjee}.
The 2d anisotropic semimetallic bandstructure can be found in systems like multi-layer VO$_2\,$-TiO$_3$ nanostructures \cite{pickett,pardo,pardo2,banerjee}, organic conductor $\alpha$-(BEDT-TTF)$_2$I$_3$ \cite{kobayashi,suzumura}, deformed graphene \cite{hasegawa,orignac,montambaux1,montambaux2}, and cold atoms trapped in an optical honeycomb lattice \cite{delplace_sdirac}.
The anisotropic nature of the spectrum manifests itself through distinctive signatures in various transport and thermodynamic properties \cite{pickett,ips-kush,ips-cd,ips-cd2,ips-kush-review}.

In this paper, we consider the  S-B-S configuration in the thin-barrier-limit, constructed with the semi-Dirac semimetal, by approximating the barrier by a Dirac delta function \cite{zagoskin,kwon_krish,asano}. In our set-up, the weak link is represented by the thin-barrier-limit of an S-B-S junction, which is defined by $ L \ll \xi $, where $ L $ is the barrier thickness and $\xi $ is the superconducting coherence length. We employ the scattering matrix approach for the associated Bogoliubov–de Gennes (BdG) Hamiltonian, which has been one of the standard techniques used extensively to determine the conductance of an N-S junction~\cite{beenakker,spin1,wsm_jj2,wsm_jj0,wsm_jj1}, with the appropriate generalization applicable for the S-B-S junctions.

The propagation direction of the quasiparticles/quasiholes is taken to be parallel/antiparallel to the $x$-axis.
We denote the transverse dimension of the junction by $W$, where $W$ is assumed to be large enough to impose periodic boundary conditions along the $z$-directiom. In the short-barrier regime, the main contribution to the Josephson current comes from the subgap Andreev states \cite{been_houten,titov-graphene,zagoskin}, because the contributions from the excited states in the continuum are suppressed by a factor of $L/\xi$. We compute the energies of the ABSs in the thin-barrier-limit, and determine the resulting Josephson current.

The paper is organized as follows. In Sec.~\ref{secham}, we describe the low-energy effective Hamiltonian of the semi-Dirac semimetal in the normal phase, and show its eigenvalues and eigenfunctions. In Sec~\ref{secsbs}, the BdG Hamiltonian, necessary to describe the S-B-S junction, is shown along with the expressions for the electronlike and holelike wavefunctions. This is followed by Sec.~\ref{secresults}, where the ABS energy values are derived and the Josephson current is computed numerically. Finally, we end with a summary and outlook in Sec.~\ref{secsum}.

\section{Semi-Dirac semimetal}
\label{secham}

\begin{figure*}[t]
\subfigure[]{\includegraphics[width = 0.5 \textwidth]{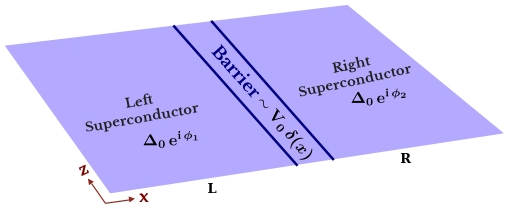}}
\hspace{2 cm}
\subfigure[]{\includegraphics[width = 0.25 \textwidth]{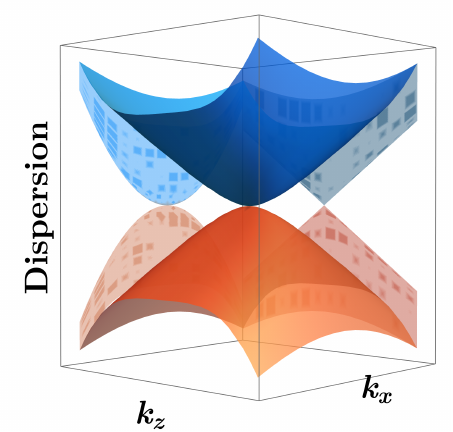}}
\caption{\label{figsetup}
Schematics of the
(a) S-B-S junction configuration;
(b) anisotropic dispersion of a semi-Dirac semimetal with a quadratic(linear)-in-momentum dependence along the $k_x$($k_z$) direction. The projections of the dispersion shown in the background
clearly demonstrate the hybrid nature.
}
\end{figure*}

An effective low-energy continuum model of the 2d anisotropic semi-Dirac semimetal, featuring a hybrid dispersion spectrum which is
linear along $k_z$ and quadratic along $k_x$, is represented by the Hamiltonian \cite{dietl,pardo,pardo2,banerjee,pickett,saha_awf,ips-kush,ips-cd,ips-kush-review} 
\begin{align}
\label{ham-continuum}
\mathcal H (\mathbf k) =\frac{\hbar^2\, k_x^2}{2\,m}\,\sigma_x +\hbar \, v\, k_z \,\sigma_z \,.
\end{align}
Here, $m$ is the effective mass parameter along the $x$-axis, $v$ is the Fermi velocity along the $z$-axis, and $\sigma_x$ and $\sigma_z$ are two of the three Pauli matrices. In order to simplify the notations, we
define the dimensionless momenta
\begin{align}
K_x =\frac{\hbar\, k_x} {p}  \text{ and } 
K_z =\frac{\hbar\, k_z } {p} \,, \text{where }
p = 2\, m\, v\,.
\end{align}
The tight-binding Hamiltonian \cite{dietl}, from which the above low-energy continuum Hamiltonian has been obtained, consists of a honeycomb lattice comprising two sublattice sites labelled as A and B (analogous to the case of graphene). Hence, there exists a pair of valleys at the two inequivalent K points/corners of the Brillouin zone, which we denote by K$_+$ and K$_-$. The Hamiltonian $ \mathcal H (\mathbf k)$ here represents the states in the vicinity of the valley located at K$_+ $.

The energy eigenvalues of $ \mathcal H (\mathbf k)$ are given by
\begin{align}
\mathcal{E}  = s \,\epsilon_0 \,\sqrt{K_x^4 + K_z^2 }\,,\quad
\epsilon_0 =\frac{p^2} {2\, m} = 2\, p\, v \,, \quad s =\pm\,,
\end{align}
as shown in Fig.~\ref{figsetup}(b).
Here, the ``$+$" and ``$-$" signs, as usual, refer to the conduction band (i.e., the upper band with the positive energy eigenvalue) and the valence band (i.e., the lower band with the negative energy eigenvalue), respectively.
Henceforth, we set $\varepsilon_0 $ to unity for uncluttering of notations, which just implies that all our energies are scaled in units of $\varepsilon_0 $. Furthermore, we will set $\hbar = 1$, except occasionally, when we retain it for the sake of clarity.

A set of two orthonormal eigenvectors is given by
\begin{align}
\label{eqev}
& \Psi^s ( \mathbf k) = 
\frac{1}{\sqrt{ K_x^4  + \left( s\, E + K_z\right)^2 }}
\left [
 s \, E + K_z \qquad K_x^2  \right]^T , \nn 
& E = \sqrt{K_x^4  +  K_z^2 } \,.
\end{align} 
For a given value of the transverse (to the propagation direction) momentum $K_z$ with $|K_z | \leq |E|$, the relation $  K_x^4  = E^2 -K_z^2  $ leads to the four solutions $K_x =\pm  \ \mathcal{K}_x  $ and $K_x =\pm \, \tilde{\mathcal K}_x  $, where $ \mathcal K_x  = \left( E^2 -K_z^2 \right)^{1/4} $ and $ \tilde{\mathcal K}_x  = i\, \mathcal K_x $. Consequently, in addition to the propagating plane wave solutions, there are also evanescent waves present~\cite{banerjee,khokhlov_nodal,Deng2020,ips-aritra}. If the Fermi energy cuts the bands at energy $E$, then for propagation along the $x$-direction, the corresponding ``right-moving'' plane waves will have the factor $e^{i\, \text{sgn}(E) \,\tilde{\mathcal K}_x \, x}$ --- this just implies that if the propagating quasiparticles are occupying the upper(lower) band, then they have a positive(negative) group velocity.

We note that the ``extra'' solutions involving the evanescent waves (i.e., the ones with the momenta $\pm \, \tilde{\mathcal K}_x$) do not exist when the propagation direction is taken to be along the $z$-axis, as the dispersion is linear-in-momentum along that direction. As a result, we expect a richer structure of the ABSs when we analyze the problem of junctions encountered for transmission along the $x$-direction, which features a nonlinear dependence on the momentum.

\section{Josephson junction}
\label{secsbs}

In order to represent the superconducting phases across the Josephson junction [cf. Fig.~\ref{figsetup}(a)], we need to define the superconducting pair potential in each region. The time-reversal operator interchanges the valleys in a graphene-like bandstructure (which has pseudospin-1/2 quasiparticles), where each valley is represented by a Hamiltonian constructed out of the Pauli matrices. Because of the valley degeneracy, it suffices to consider one of the two possible sets, thus leading to a $4\times 4$ BdG Hamiltonian.
Adopting a homogeneous approximation, we follow the construction in Refs.~\cite{beenakker,krish-moitri}, such that the pair potential can be modelled as
\begin{align}
\Delta (x) =\begin{cases} 
\Delta_0\,e^{i\,\phi_1 }\,\sigma_0
 &\text{ for }  x < 0   \\
\Delta_0\,e^{i\,(\phi_1+\phi ) } \,\sigma_0  &\text{ for } x > 0
\end{cases}, \quad
\sigma_0 = {\mathbb{1}}_{2\times 2} \,,
\end{align}
representing BCS-like Cooper pairing in the spin-singlet s-wave channel. The phases of the superconducting order parameter in the two regions are given by $\phi_1$ and $\phi_1 +\phi $, respectively, such that the phase difference is $\phi$. Since the final expressions for the ABSs and the Josephson current depend  only on $\phi$, we set $\phi_1= 0 $, without any loss of generality.
Due to the presence of the Delta function potential barrier between the two superconductors, we need to consider the potential energy function
\begin{align}
V(x)
= V_0 \,\delta(x) \,.
\end{align} 

In order to contrast our scenario with the linear-in-momentum dispersion cases, a few important points need to be remembered here. Although the thin-barrier-limit is equivalent to a Dirac delta potential, we do not have any constraint on the derivatives of the wavefunction across the junctions when the dispersion is linear, implying that the
standard delta function potential approximation for thin barriers cannot be taken from the outset \cite{krish-moitri}. For those situations, we need to start with a finite normal state region (rather than a Dirac delta function), obtain the equations from the boundary conditions at the S-B and B-S junctions, and finally impose the appropriate limits while computing the final solutions \cite{titov-graphene,krish-moitri, sumathi-udit,debabrata-krish,ips_jj_rsw}. However, for a quadratic dispersion, we can use the Dirac delta approximation from the start, because here we have a constraint on the first order derivatives (with respect to the position coordinate along the propagation direction) of the wavefunction across the junction, analogous to the tunneling problem involving a Schrödinger particle.

The BdG Hamiltonian can be constructed as \cite{beenakker, krish-moitri}
\begin{widetext}
\begin{align}
& H = \sum_{\mathbf r}
\Psi^\dagger (\mathbf r) \,H_{\text{BdG}}(\mathbf r) \,
\Psi (\mathbf r), \quad
\Psi(\mathbf r) = \begin{bmatrix}
c_{A+} (\mathbf r) & &  c_{B+} (\mathbf r)  & &
c_{A-}^\dagger (\mathbf r) & & -  c_{B-} ^\dagger(\mathbf r)
\end{bmatrix}^T, \nn
& H_{\text{BdG}} (\mathbf r) =
\begin{bmatrix}
\mathcal{H} (\mathbf  K \rightarrow \boldsymbol -i \,\nabla_{\mathbf r} ) 
-E_F  + V(x)  & & \Delta(x)  \\  \\
 \Delta^\dagger(x) & & E_F - V(x) -\mathcal{H} (\mathbf  K 
 \rightarrow \boldsymbol -i \,\nabla_{\mathbf r} )  \\  
\end{bmatrix},
\label{eq_bdg}
\end{align}
\end{widetext}
where $\mathbf r=(x, \, z)$ is the position vector, and the indices $\pm$ on the fermionic operators label the two valleys.

Here we demarcate the left superconducting region as ``L'' and the right superconducting region as ``R'', with the delta function barrier being the weak link in the middle. The electronlike and holelike BdG quasiparticles are obtained from the eigenvalue equation
\begin{align}
H_{\text{BdG}} ({\mathbf r}) \,\psi_{\mathbf K} (\mathbf r) 
= \varepsilon \, \psi_{\mathbf K} (\mathbf r) \,.
\end{align}
If $ \psi_N (\mathbf K )$ denotes an eigenfunction of $\mathcal{H} (\mathbf  K)$, then the electronlike and holelike eigenfunctions of $ H_{\text{BdG}} ( \mathbf r ) $ are of the forms $\psi_e (\mathbf K, \varphi)\, e^{i\,\mathbf K\cdot \mathbf r}$ and $\psi_h (\mathbf K, \varphi)\, e^{i\,\mathbf K\cdot \mathbf r}$, respectively, where~\cite{timm}
\begin{align}
\label{eqelechole}
& \psi_e (\mathbf K, \varphi) = \begin{bmatrix}
\psi_N (\mathbf K) & & & & \frac{ \left( \epsilon -\Omega \right) \, e^{-i\, \varphi }} {\Delta_0}\,
 \psi_N (\mathbf K)
\end{bmatrix} \text{ and }
\nn &
\psi_h (\mathbf K, \varphi) = \begin{bmatrix}
\psi_N (\mathbf K) & & & & \frac{ \left( \epsilon + \Omega \right) \, e^{-i\, \varphi }} {\Delta_0}\,
 \psi_N (\mathbf K)
\end{bmatrix} .
\end{align}
Here, 
\begin{align}
\Omega = i \,\sqrt{ \Delta_0^2 - \epsilon^2 } \,,
\end{align}
and $  \varphi $ represents the phase of the superconducting order parameter.

Let us define the variable
\begin{align}
\beta =\arccos(\varepsilon /\Delta_0)\,,
\end{align}
which will be useful in the expressions that follow.
Using Eqs.~\eqref{eqev} and \eqref{eqelechole}, let us now spell out the form of the eigen wavefunction
\begin{align}
 \Psi (\mathbf r, K_z ) = \psi_{L} (\mathbf r, K_z) \,\Theta(-x)
  +  \psi_{R} (\mathbf r , K_z) \,\Theta(x) \,,
\end{align}  
expressed
in a piecewise manner for the two regions, setting the chemical potential at $E_F  > 0$. We assume that\footnote{The condition $\Delta_0 \ll E_F $ ensures that the mean-field approximation, applicable for using the BdG formalism, is valid. The second condition $ (V_0-E_F ) \gg E_F  $ arises because we are focussing on the short-barrier regime.} $V_0 \gg E_F   \gg \Delta_0 $ and $(V_0-E_F ) \gg E_F  $. Since the translation symmetry is broken along the $x$-axis, $K_x$ is not conserved, whereas the transverse momentum component $K_z $ remains unchanged across the junction.

\begin{figure*}[t]
\centering
\subfigure[]{\includegraphics[width=0.4 \textwidth]{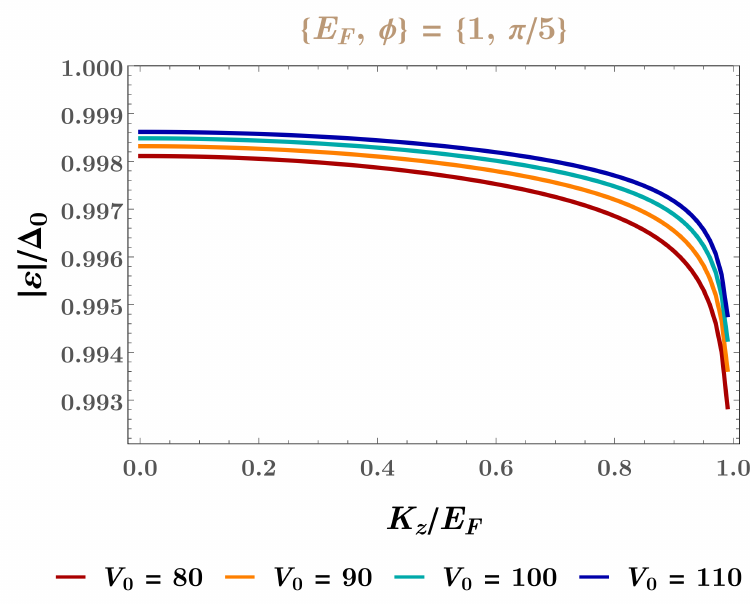}}\hspace{2 cm}
\subfigure[]{\includegraphics[width=0.4 \textwidth]{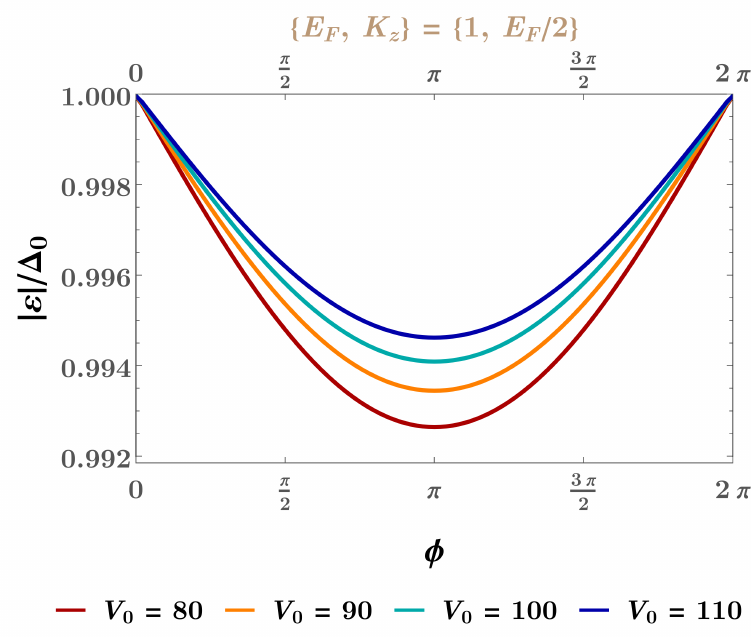}}
\caption{
Behaviour of $| \varepsilon | $ as a function of (a) $K_z $ and (b) $ \phi $, for some representative values of $V_0$ (shown in the plotlegends), with $E_F $ set to unity.
\label{fige}}
\end{figure*}

\begin{enumerate}

\item In the left superconductor region, we need to construct a linear combination of the form
\begin{align}
\psi_{L}  ( \mathbf r, K_z ) &= 
a_l \, \psi_e ( -\kappa_x^e, K_z, 0 )\, e^{ - i\,\kappa_x^e \, x +  K_z \, z }
\nn & \quad 
+ b_l \, \psi_h ( -\kappa_x^h, K_z,0 )\, e^{ - i\,\kappa_x^h \, x +  K_z \, z }
\nn & \quad 
+ c_l \, \psi_e ( -\tilde \kappa_x^e, K_z,0 )\, e^{ - i\,\tilde \kappa_x^e \, x +  K_z \, z }
\nn & \quad 
+ d_l \, \psi_h ( \tilde \kappa_x^h, K_z,0 )\, e^{ i\,\tilde \kappa_x^h \, x +  K_z \, z }
\,,
\end{align}
where
\begin{widetext}
\begin{align}
\psi_e ( K_x, K_z, \varphi )
 & \simeq 
\left [ e^{i\, \beta } \left( E_F  +K_z\right)
\qquad e^{i \,\beta } \,K_x^2
\qquad \qquad
 e^{-i \,\varphi } \left( E_F  +K_z  \right)
 \qquad e^{-i \,\varphi } \, K_x^2
 \right ]^T  , \nn
\psi_h ( K_x, K_z, \varphi )
&  \simeq 
\left [ 
K_z- E_F   \qquad \qquad \quad
K_x^2 \qquad \qquad 
e^{i \,\beta -i \,\varphi } \left(K_z- E_F  \right)
\qquad
e^{i \,\beta -i\, \varphi }\, K_x^2
 \right ]^T  ,
\end{align}
\begin{align}
& \kappa_x^e \simeq 
k_{\rm{mod} }
+ i \,\kappa \,,\quad
 k_{\rm{mod} } \simeq 
\left(  E_F ^2  - K_z^2 \right)^{1/4} \,,
\quad 
\kappa  \simeq  \rm{Im} \Big[
\left \lbrace   \left( E_F  + i\, \Delta_0 \right)^2  
- K_z^2 \right \rbrace ^{1/4} \Big]\,, \nn
& \kappa_x^h \simeq 
 -k_{\rm{mod} }
+ i \,\tilde \kappa \,,\quad
\tilde \kappa  \simeq  \rm{Im} \Big[
\left \lbrace   \left( E_F  - i\, \Delta_0 \right)^2  
- K_z^2 \right \rbrace ^{1/4} \Big]\,, \quad
\tilde \kappa_x^e  = i\, \kappa_x^e \,,\quad
\tilde \kappa_x^h  = i\, \kappa_x^h \,.
\end{align}
\end{widetext}

\item In the right superconductor region, the wavefunction localizing at the
interface is described by the linear combination (cf. chaper 5 of Ref.~\cite{asano})
\begin{align}
\psi_{R}  ( \mathbf r, K_z ) &=
a_r \, \psi_e ( \kappa_x^e, K_z, \phi )\, e^{  i\,\kappa_x^e \, x +  K_z \, z }
\nn & \quad 
+ b_r \, \psi_h ( \kappa_x^h, K_z, \phi )\, e^{  i\,\kappa_x^h \, x +  K_z \, z }
\nn & \quad 
+ c_r \, \psi_e ( \tilde \kappa_x^e, K_z, \phi )\, e^{  i\,\tilde \kappa_x^e \, x +  K_z \, z }
\nn & \quad 
+ d_r \, \psi_h ( -\tilde \kappa_x^h, K_z, \phi )\, e^{ -i\,\tilde \kappa_x^h \, x +  K_z \, z }
\,.
\end{align}


\end{enumerate}

Imposing the continuity of the wavefunction and the constraint on its first order derivatives (with respect to $x$) across the junction, located at $x =0$, we get the following equations:
\begin{align}
\label{eqbdy}
& \psi_{L}  (0, z, K_z) = \psi_{R}  (0, z, K_z) \text{ and } 
\nn &
\partial_x \psi_{R}  (x, z, K_z) \big \vert_{x=0}
- \partial_x \psi_{L}  (x, z, K_z) \big \vert_{x=0}
= V_0 \, \psi_{L}  (0, z, K_z) \,.
\end{align}
From the four components of the BdG wavefunction, we get
$2\times 4 = 8 $ linear homogeneous equations in the $ 8 $ unknown variables $ \left( a_{l}, \, b_{l},\, 
c_l, \, d_l ,\, a_{r}, \, b_{r},\, 
c_r, \, d_r \right)$, which constitute the coefficients of the piecewise-defined wavefunction. In the resulting equations, the overall $ z $-independent factors of $e^{ i \, K_z\, z}$ cancel out.
Let $ \mathcal M$ denote the $ 8 \times 8  $ matrix constructed out of the coefficients of the 8 variables. For the equations to be consistent, we need to impose the condition $\text{det}\, \mathcal M = 0 $. This helps us determine the energy eigenvalues of the subgap ABSs, which are localized near the junction, since they decay exponentially with the distance from the weak link into the superconducting region on either side.

\section{Results}
\label{secresults}

\begin{figure*}[t]
\centering
\subfigure[]{\includegraphics[width=0.35 \textwidth]{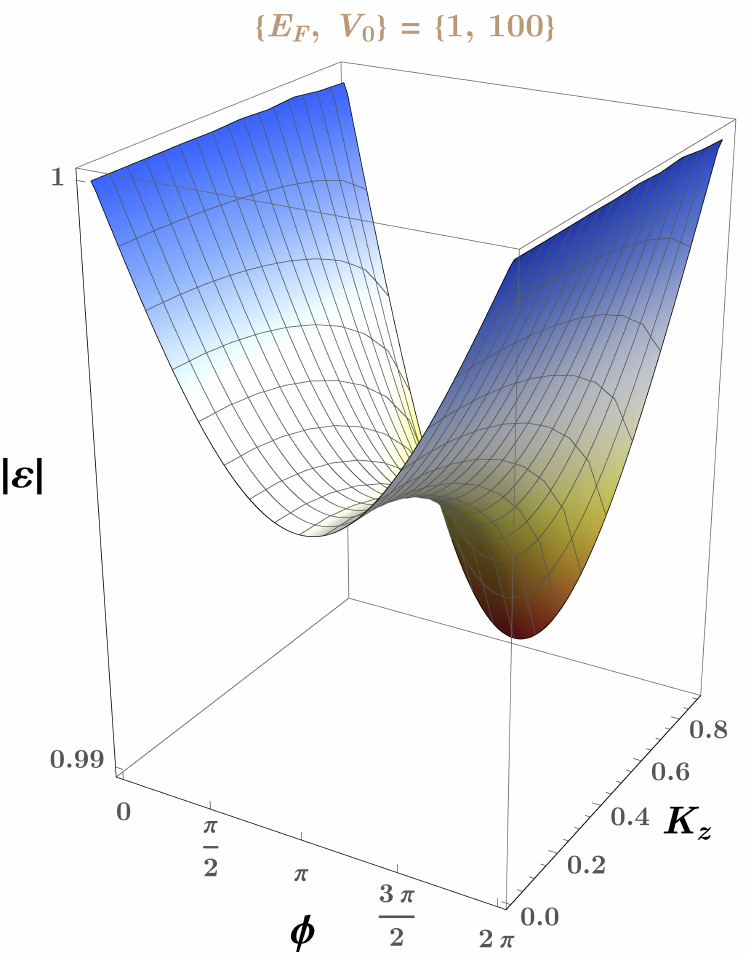}}\hspace{2 cm}
\subfigure[]{\includegraphics[width=0.475 \textwidth]{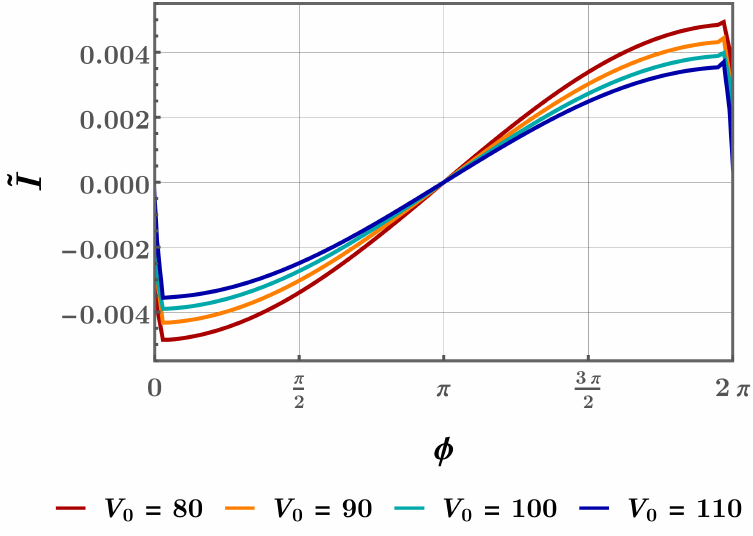}}
\caption{
(a) Magnitude of the energy $\varepsilon $ of the Andreev bound states against the $ \phi \,$-$K_z $ plane, for $E_F  = 1$ and $V_0 = 100$. (b) The behaviour of the total Josephson current ($\propto \tilde I $), in arbitrary units, as a function of $ \phi $, obtained at $E_F  = 1$ and $k_B \, T = 0.005\, \Delta_0 $. We have used four values of $V_0$ as shown in the plotlegends.
\label{figj}}
\end{figure*}

In order to reduce the complexity of the computations to the determinant of a lower dimensional matrix, we first eliminate four of the eight unknown variables by using four of the eight linear homogeneous equations. Specifically, in our calculations, we first solve for $ \left( a_{l}, \, b_{l},\, c_l, \, d_l \right)$ in terms of $ \left( a_{r}, \, b_{r},\, c_r, \, d_r \right)$, using $\psi_{L}  (0, z, K_z) = \psi_{R}  (0, z, K_z) $. These solutions are plugged into the constraint
$$\partial_x \psi_{R}  (x, z, K_z) \big \vert_{x=0}
- \partial_x \psi_{L}  (x, z, K_z) \big \vert_{x=0}
= V_0 \, \psi_{L}  (0, z, K_z)\,, $$ such that $ \left( a_{l}, \, b_{l},\, c_l, \, d_l \right)$ no longer appear in the equations resulting from the operation. Now we construct the $ 4 \times 4  $ matrix $ \tilde {\mathcal M}$ out of the coefficients of the four variables, viz., $ \left( a_{r}, \, b_{r},\, c_r, \, d_r \right)$. The explicit form of this matrix is very long and, therefore, we write the 4 columns one by one:
\begin{widetext}
\begin{align}
c_1 = 
\begin{bmatrix}
 e^{i  \, \beta } \left(E_F ^2-K_z^2\right) 
 \left [
 e^{i \, \phi } \left \lbrace
 E_F  \left(-z \, V_0
 + (1+i) (1+ \left( 1+i \right) z) \kappa _x\right)+K_z \left(-z \, V_0+2 \, i \left (z+1 \right ) \kappa _x\right)
 \right \rbrace
 - \left( 1+i \right) \kappa _x \left \lbrace E_F + \left( 1+i \right) K_z \right \rbrace
 \right ] \\ \\
 e^{i  \, \beta } \kappa _x^2 \left(E_F + K_z\right) 
 \left[     e^{i  \,\phi } 
 \left \lbrace
 E_F  \left(-z \, V_0+ \left( 1+i \right) (1+ \left( 1+i \right) z) \kappa _x\right)+K_z \left(z \, V_0-2 \, i \left (z+1 \right ) \kappa _x\right) \right \rbrace
 - \left( 1+i \right) E_F  \,\kappa _x
 + 2 \, i \,K_z \,\kappa _x \right ] \\ \\
 \left(E_F ^2-K_z^2\right) \left [
 E_F  \left \lbrace -z \, V_0+ \left( 1+i \right) \kappa _x \left(\left (z+1 \right ) e^{i \, \phi }
 + i \, z-1\right)\right \rbrace 
 + K_z \left \lbrace -z \, V_0+2 \, i \left(-1+e^{i \,(2\, \beta +\phi )}\right) \kappa _x \right \rbrace 
 \right ] \\ \\
 \kappa _x^2 \left(E_F +K_z\right) 
 \left [
 E_F  \left \lbrace 
 -z \, V_0 + \left( 1+i \right) \kappa _x \left(\left (z+1 \right ) e^{i \, \phi }+ i \, z-1\right)
 \right \rbrace 
 + K_z \left \lbrace z \, V_0-2 \, i \left(-1+e^{i\, (2\, \beta +\phi )}\right) 
 \kappa _x \right \rbrace  \right ] 
\end{bmatrix},
\end{align}
\begin{align}
c_2 = 
\begin{bmatrix}
 \left( E_F ^2-K_z^2\right) \left[
 e^{i \,\phi } \left \lbrace
 V_0\,  z  \left( E_F -K_z\right)+ \left (1+i \right )  E_F  
 \left (z-i \right ) \kappa _x + 2\, i\, K_z \,\kappa _x 
 \right \rbrace
 - \left (1-i \right ) \left (z+1 \right ) \kappa _x 
 \left \lbrace E_F - \left (1-i \right ) K_z \right \rbrace
 \right ] \\
 \\
 \kappa _x^2 \left( E_F -K_z\right) \left [
 e^{i \, \phi } \left \lbrace
 -V_0\,  z  \left( E_F +K_z\right)
 + \left (1+i \right ) \kappa _x \left( \left (1+i \right ) K_z
 - E_F  \left (z-i \right )
 \right) \right \rbrace 
+ \left (1-i \right )  \left ( z+1 \right ) \kappa _x 
\left \lbrace  E_F + \left (1-i \right ) K_z \right \rbrace 
 \right ] \\
 \\
 e^{i \, \beta } \left( E_F ^2-K_z^2 \right) 
 \left [ 
 E_F  \left \lbrace 
 V_0\,  z + \left (1+i \right ) \kappa _x \left (
   \left (1+i \right ) z -i \,e^{i \, \phi} + i \right )
 \right \rbrace 
 + K_z \left \lbrace -V_0\,  z + 2 \, i \, \kappa _x \left(e^{i \, \phi } - z -1\right)
 \right \rbrace
 \right ] \\
 \\
 e^{i \, \beta } \kappa _x^2 
 \left( E_F -K_z\right) \left [
 K_z \left \lbrace -V_0\,  z + 2 \, i \, \kappa _x \left( e^{i\, \phi } - z-1\right)
 \right \rbrace 
 - E_F  \left \lbrace
 V_0\,  z + \left (1+i \right ) \kappa _x 
 \left ( \left (1+i \right ) z -i \,e^{i \, \phi} + i \right )
 \right \rbrace 
 \right ] 
\end{bmatrix},
\end{align}
\begin{align}
c_3 =
\begin{bmatrix}
 e^{i \left (\beta +\phi \right )} 
 \left( E_F ^2-K_z^2\right) \left[
 -z\, K_z \left(V_0 + 2 \,\kappa _x \right)
 + E_F  \left \lbrace  - V_0 \, z - \left (1-i \right ) \kappa _x 
 \left ( \left (1+i \right ) z - e^{-i\,\phi } +1  \right )\right \rbrace
 \right ] \\ \\
 e^{i \,\beta } \,\kappa _x^2 \left( E_F +K_z\right) 
 \left [
   e^{i  \,\phi } \left \lbrace
 -z \,K_z \left(V_0 + 2 \,\kappa _x\right)
 + E_F  \left( V_0\, z+ \left (1-i \right ) \kappa _x
 + 2\, z \, \kappa _x\right)\right \rbrace
  - \left (1-i \right )  E_F \,  \kappa _x
 \right ] \\ \\
 \left( E_F ^2-K_z^2\right) 
 \left [-z \,K_z \left(V_0 + 2 \, \kappa _x\right)
 - E_F  \left \lbrace  V_0\, z+ \left (1+i \right ) \kappa _x 
 \left ( -i \,e^{i\,\phi}  \left (z+1 \right ) + z+i  \right )
 \right \rbrace \right ]\\ \\ 
 \kappa _x^2 \left( E_F +K_z\right) \left [
 -z \, K_z \left(V_0 + 2\, \kappa _x\right)
 + E_F  \left \lbrace  V_0\, z+ \left (1+i \right ) \kappa _x 
\left ( -i \, e^{i\,\phi} \left (z+1 \right )   +z+i \right )
 \right \rbrace\right ]
\end{bmatrix},
\end{align}
\begin{align}
c_4 =
\begin{bmatrix}
 \left( E_F^2-K_z^2\right) \left [  e^{i  \,\phi } \left \lbrace
 - z\, K_z  
 \left(V_0+ 2 \, \kappa _x\right)
 + E_F \left (
 V_0  \, z+ \left (1-i \right ) \left (z-i \right ) \kappa _x \right) \right \rbrace
 + \left (1+i \right )  E_F \left  (z+1 \right ) \kappa _x
 \right] \\ \\
 \kappa _x^2 \left( E_F-K_z\right) \left [ 
 e^{i  \,\phi } \left \lbrace
  z\, K_z  \left(V_0 + 2 \, \kappa _x\right)
 + E_F \left(V_0 \, z
 + \left (1-i \right ) \left  (z-i \right ) \kappa _x\right)\right \rbrace
 + \left (1+i \right )  \left  (z+1 \right )  E_F\, \kappa _x
 \right ] \\ \\
 -e^{i \,\beta } 
 \left ( E_F^2-K_z^2 \right) 
 \left [
  -z \,E_F\, V_0 + \left (1+i \right )  E_F\, \kappa _x 
 \left \lbrace  \left (i-1 \right ) z+  e^{i  \,\phi }-1 \right \rbrace 
 + z\, K_z  \left(V_0+ 2 \, \kappa _x\right)\right ] \\ \\
 e^{i \,\beta } \kappa _x^2 \left( E_F-K_z\right) \left[
  z\, K_z  \left(V_0 + 2 \, \kappa _x\right)
 + E_F \left \lbrace V_0 \,z- \left (1+i \right ) \kappa _x 
 \left( \left (i-1 \right ) z+  e^{i  \,\phi }-1 \right)
 \right \rbrace \right ]
\end{bmatrix},
\end{align}
\end{widetext}
such that
$$ \tilde {\mathcal M} = e^{-i\,\phi} \,\begin{bmatrix}
c_1 \quad c_2 \quad c_3 \quad c_4
\end{bmatrix} \,.$$
Here, $z =  e^{2\, i\,\beta } - 1 $.
The consistency condition reduces to $ \text{det}\,\tilde {\mathcal M}= 0 $. The equation resulting from this vanishing determinant gives a lower order polynomial equation in $e^{i\, \beta}$ (compared to the one obtained from $\text{det}\, \mathcal M = 0 $) and, hence, is easier to solve. In fact, we obtain a quartic equation in the variable $ z$, as shown below:
\begin{widetext}
\begin{align}
\label{eqdet}
& z^4 \left(K_z^2-E_F ^2\right)^2
 \left(V_0 + 2 \,{\mathcal K}_x\right)^2 
\left(V_0^2 + 4 \,{\mathcal K}_x^2\right)
-8 \, z^3 \left(K_z^2- E_F ^2 \right)  {\mathcal K}_x^2  
\left ( \cos \phi -1 \right ) 
\left(2 \,{\mathcal K}_x + V_0\right) 
\left [ 2 \left(K_z^2-E_F ^2 \right) {\mathcal K}_x  + V_0 \, K_z^2 \right ]
\nn
& +
16 \,z^2 \, {\mathcal K}_x^2 \,\sin ^2 \Big (\frac{\phi }{2}\Big  ) 
\left [
-E_F ^2 \, K_z^2 \left(V_0 + 2 \, \kappa_x \right) \left(V_0 + 4\, \kappa_x  \right)
+ 2\, E_F ^4 \,\kappa_x  \left(V_0 + 3 \, \kappa_x  \right)
+ K_z^4 \left(V_0 + 2 \, \kappa_x  \right)^2
-2\, E_F ^4 \, {\mathcal K}_x^2  \cos \phi 
\right ]
\nn &
+ 128\,z \,E_F ^2 \,{\mathcal K}_x^4 \, \sin ^4\left(\frac{\phi }{2}\right)
+ 64\,E_F ^2\, {\mathcal K}_x^4 \, \sin ^4 \Big (\frac{\phi }{2} \Big  ) = 0\,.
\end{align}
\end{widetext}
The order of this polynomial equation (whose roots we need to determine) is the same as what we found for the case of the linearly dispersing Rarita-Schwinger-Weyl (RSW) semimetal~\cite{ips_jj_rsw}, featuring four bands (rather than two).

The coefficients of various powers of $z$ in Eq.~\eqref{eqdet} are all real and, hence, we can analyze the nature of the roots by applying the general criteria applicable for a real quartic polynomial equation. First we obtain the associated depressed quartic, which takes the form:
$$ z^4 + q\, z^2 + \rho \, z + \gamma = 0 \,. $$
Let $\mathcal D$ be the discriminant of the depressed polynomial. For a given set of values for $E_F $, $V_0$, and $K_z$, we compute numerically the values of  $\mathcal D$, $q$, and $\gamma $. For each case, we find that $ \mathcal D >0$.
This means that if $q>0$ or $4 \,\gamma -q^2 >0 $, we get a pair of complex conjugate roots (i.e., we get four complex roots which are of the form $z_1$, $z_1^*$, $z_2$, and $z_2^*$).
This is what we find from our numerical simulations. However, since $z =   e^{2\, i\,\beta } - 1 $, an admissible
solution must satisfy the condition $ |  z + 1 |  = 1 $. From our numerical data, we
find that only one set of complex conjugate roots, out of the pair, fall in the allowed
category. This is in contrast with the RSW case \cite{ips_jj_rsw}, where we get, in general, four distinct solutions for $|\varepsilon|$.

From the above analysis, we arrive at the conclusion that the energies of the subgap states appear as the pairs $\pm |\varepsilon| $ for a doubly degenerate value of $ |\varepsilon| $. Therefore, for each value of $|\varepsilon|$, we get a total of four ABSs --- two with the value $|\varepsilon|$ and two with the value $ - |\varepsilon|$. To demonstrate the results, we include some representative plots, and all of these are obtained by setting $E_F =1$. In Fig.~\ref{fige}, we show the behaviour of $|\varepsilon|$ as a function of (a) $ K_z $ (with a fixed value of $\phi $), and (b) $\phi $ (with a fixed value of $  K_z $), for some representative values of $V_0$. The bound state energies are periodic in $\phi$ with period $2\pi $. They are, in fact, functions of $\cos \phi $, as is evident from Eq.~\eqref{eqdet}. This dependence is reflected in the nature of the curves in Fig.~\ref{fige}(b). Fig.~\ref{figj}(a) illustrates the variation of $ |\varepsilon| $-values against the $\phi \,$-$K_z$ plane and, hence, shows the dependence of the bound state energies on both these variables in a combined way.

The Josephson current density across the junction at a temperature $T$ is given by \cite{zagoskin,titov-graphene}
\begin{align}
I_J( \phi) = -\frac{2\, e}{\hbar}\,\frac{W} {2\,\pi} \sum_{n=1} ^ 4
\int  dK_z \, \frac{ \partial \epsilon_n}
{ \partial  \phi } \,f(\epsilon_n) \,,
\end{align}
where $\epsilon_n$ labels the energy values of the four Andreev bound states, and $f(\zeta) = 1 / \left(  1+e^{\frac{\zeta}{k_B\, T}} \right)$ is the Fermi-Dirac distribution function. Fig.~\ref{figj}(b) shows the behaviour of $I_J$ as a function of $\phi $, scaled by appropriate numbers/variables (this scaled quantity being denoted as $\tilde I $), for four values of $V_0$.

\section{Summary and outlook}
\label{secsum}

In this paper, we have have computed the characteristics of the emergent ABSs, and the resulting Josephson current, in a Josephson junction built with a 2d semi-Dirac semimetal. The junction is taken to be perpendicular to the quadratically dispersing direction in this anisotropic material, which features a hybrid dispersion.
The weak link between the two superconducting regions is modelled by a Dirac delta function potential. 
Using the BdG formalism, we have determined the wavefunction localizing at the junction for $ |\varepsilon | \ll \Delta_0 $. This requires a piecewise continuous definition --- the usual procedure adopted for solving reflection and transmission problems in quantum mechanics involving junction configurations. 
Demanding consistency of the equations obtained from matching the boundary conditions at the junction, we arrive at a quartic polynomial equation in an appropriate variable variable (related to the modulus of the ABS energy $\varepsilon$), obtained from the vanishing of the relevant determinant. 
The physically admissible roots of this quartic give the discrete energy spectrum $\varepsilon $ of the ABSs. Although a closed form solution cannot be found, due to the fourth order polynomial involved, we have deduced the nature of the roots from our numerical data. We have found that the values of
$| \varepsilon | $ are doubly degenerate, leading to four ABSs --- two with energy $ | \varepsilon| $ and two with energy $ - |\varepsilon| $. We have also shown that the solutions depend on the phase difference ($\phi$) between the two superconducting regions via functions of $\cos \phi $.

Our main finding is that, because of the quadratic dispersion, we need to include the ``evanescent'' wave solutions, while defining the wavefunctions in each region. This is in stark contrast with the cases where the propagation axis is along a direction of linear-in-momentum dispersion \cite{titov-graphene,krish-moitri,sumathi-udit,debabrata-krish,debabrata,ips_jj_rsw}. The existence of the extra solutions results in a higher order polynomial equation to be solved, compared to the analogous linearly dispersing cases \cite{titov-graphene,krish-moitri,sumathi-udit,debabrata-krish,debabrata} with the same number of bands involved. In fact, comparing our results with the isotropic four-band RSW system studied earlier \cite{ips_jj_rsw}, we find that the order of the polynomial, for propagation along the quadratic-in-momentum dispersion in a two-band system, is the same as that in a linearly dispersing four-band system. 

In this paper, we have considered the Josephson current in the thin-barrier-limit, which allows one to model the barrier as a Dirac delta function. One can consider a finite barrier instead, but the computations will be significantly challenging. In such cases, we have a finite normal state region sandwiched between two superconducting regions, and we need to determine the piecewise continuous wavefunctions from the boundary-condition-matching at the S-B and B-S junctions
at $ x=0 $ and $ x =\ell$, where $\ell$ is the length of the barrier along the propagation direction.
Furthermore, barriers of more generic shapes (rather than the simple rectangular form) can be considered, which might be realized in futuristic experimental set-ups. Yet another interesting barrier-configuration is to consider scenarios where the anisotropic dispersion is rotated about the propagation-axis across the junction(s), as considered in Refs.~\cite{tilted_jn,ips_jns}.

In the future, it will be worthwhile to study the Josephson effect in 3d multi-Weyl semimetals \cite{debabrata-krish,debabrata,Deng2020,ips-aritra}, by considering the Josephson junction aligned perpendicular to one of the nonlinearly dispersing directions.
Another interesting avenue is to introduce a tilt in the Hamiltonians of various nodal-point semimetals~\cite{debabrata}, and investigate the resulting ABSs. Last but not the least, analysis of Josephson junctions built with isotropic Luttinger
semimetals, harbouring quadratic band crossing points~\cite{Boettcher,ips-qbt-sc}, is left for future work.

\section*{Acknowledgments}
We thank Angad Singh for a careful reading of the manuscript.

\bibliography{biblio_jj}
\end{document}